\documentclass[a4paper,12pt]{article}
\def\al{\alpha}
\def\be{\beta}
\def\ga{\gamma} \def\Ga{\Gamma}
\def\ep{\epsilon}
\def\lam{\lambda}

\def\calD{{\cal D}}  \def\calF{{\cal F}}
  
  \def\calL{{\cal L}}

\def\del        {  \partial  }
\def\half       {  {1\over 2}  }
\def\trace      {  \mbox{Tr}  }
\def\Tr         { {\rm Tr} }
\def\ie         {  {\it i.e.}      }

\def\comma          {\, ,}
\def\period         {\, .}
\def\lsim    {\lower .65ex \hbox{\ $\stackrel{<}{\sim}$\ } }
\def\gsim    {\lower .65ex \hbox{\ $\stackrel{>}{\sim}$\ } }
\def\com#1#2   { \left[#1, #2\right]} 
\def\acom#1#2  {\left\{ #1,#2\right\}}
\def\bra#1     {\langle #1 |}
\def\ket#1     {| #1 \rangle}
\def\slash#1{{\ooalign{\hfil/\hfil\crcr$#1$}}} 
\def\vecii#1#2      {  \left(\begin{array}{c}#1\\#2\end{array}\right)  }
\def\veciii#1#2#3   {  \left(\begin{array}{c}#1\\#2\\#3\end{array}
                     \right)  }
\def\veciv#1#2#3#4  {  \left(\begin{array}{c}#1\\#2\\#3\\#4
                                 \end{array}\right)  }
\def\vecfv#1#2#3#4#5 {  \left(\begin{array}{c}#1\\#2\\#3\\#4\\#5
                                 \end{array}\right)  }

\def\matrixii#1#2#3#4            {  \left(\begin{array}{cc}#1&#2\\#3&#4
                                       \end{array}\right) }
\def\matrixiii#1#2#3#4#5#6#7#8#9 {  \left(\begin{array}{ccc}#1&#2&#3\\
                                     #4&#5&#6\\#7&#8&#9\end{array}
                               \right)  }
\def\mativ#1#2#3#4               {  \left(\begin{array}{cccc}
                                       #1\\#2\\#3\\#4\end{array}\right) }
\def\matv#1#2#3#4#5              {  \left(\begin{array}{ccccc}
                                     #1\\#2\\#3\\#4\\#5\end{array}
                              \right)  }
\def\eqabegin         {  \begin{eqnarray}  }
\def\eqaend           {  \end{eqnarray}  }
\def\nn               {  \nonumber  }
\def\bracetwo#1#2     {  \left\{ \begin{array}{l} #1 \\ #2 \end{array}
                         \right.  }
\def\bracetwocases#1#2#3#4  {   \left\{ \begin{array}{ll} #1 &
                                 \qquad #2 \\
                                 #3 & \qquad #4 \end{array} \right.  }
\def\bracebegin#1     {  \left\{ \begin{array}{#1}   }
\def\braceend         {  \end{array}\right.   }
\def\parn              {  \par\noindent }

\def\parmedskip        {  \par\medskip  }
\def\parsmallskip      {  \par\smallskip  }
\def\parbigskipn        {  \par\bigskip\noindent  }
\def\parmedskipn        {  \par\medskip\noindent  }
\def\parsmallskipn      {  \par\smallskip\noindent  }

\def\parag#1           {\paragraph{#1} \mbox{ }\parmedskip\noindent}
\def\msection#1      {  \begin{center} \section{#1} \end{center}   }
\def\nsection#1      {  \let\boldface\bf \def\bf{} \section{#1}
                           \let\bf\boldface   }
\def\mnsection#1     {  \begin{center} \nsection{#1} \end{center}  }
\def\capsection#1    {  \let\boldface\bf \def\bf{\sc} \section{#1}
                           \let\bf\boldface   }
\def\mcapsection#1   {  \begin{center} \capsection{#1} \end{center} }

\def\sectionnumbering { \setcounter{equation}{0}
         \renewcommand{\theequation}{\arabic{section}.\arabic{equation}}}

\newcommand{\nullify}[1]{}
\def\papertitlepage{\baselineskip 3.5ex \thispagestyle{empty}}
\def\Title#1{\baselineskip 1cm \vspace{1.5cm}\begin{center}
 {\Large\bf #1} \end{center} 
\vspace{0.5cm}}
\def\Authors#1{\begin{center} {\it #1} \end{center}}
\def\Abstract{\vspace{1.0cm}\begin{center} {\large\bf Abstract} 
           \end{center} \par\bigskip}
\def\Komabanumber#1#2#3{\hfill \begin{minipage}{4.2cm} UT-Komaba #1
              \parn #2 
              \parn #3 \end{minipage}}
\renewcommand{\thefootnote}{\fnsymbol{footnote}}
\renewenvironment{thebibliography}{\pagebreak[3]\par\vspace{0.6em}
\begin{flushleft}{\large \bf References}\end{flushleft}
\vspace{-1.0em}

\begin{enumerate}\if@twocolumn\baselineskip=0.6em\itemsep -0.2em
\else\itemsep -0.2em\fi\labelsep 0.1em}{\end{enumerate}}

\def\Gatil{\widetilde{\Ga}}

\def\Gazero{\Ga^{(0)}}

\def\rslash{\slash{r} }

\def\thdot{\dot{\theta}}

\def\dzep{\delta^{(0)}_\ep}
\def\dzlam{\delta^{(0)}_\lam}
\def\Gazero{\Ga^{(0)}}
\def\thdot{\dot{\theta}}
\def\thddot{\ddot{\theta}}
\def\thddot{\ddot{\theta}}

\def\thd#1{\theta^{(#1)}}
\def\Lemma#1{\parbigskipn {\it Lemma #1 :} \quad }
\def\Proof{\parmedskipn {\it Proof:} \quad }

\def\dldal#1{{\del \calL \over \del #1_\al}}

\def\qed{\quad {\it Q.E.D}}
\newcommand{\barr}{\begin{array}}
\newcommand{\earr}{\end{array}}

\def\and{&}
\usepackage{graphicx}
\usepackage{latexsym}
\usepackage{amsmath}
\usepackage{amssymb}

\renewenvironment{thebibliography}{\pagebreak[3]\par\vspace{0.6em}
\begin{flushleft}{\large \bf References}\end{flushleft}
\vspace{-1.0em}

\begin{enumerate}\if@twocolumn\baselineskip=0.6em\itemsep -0.2em
\else\itemsep -0.2em\fi\labelsep 0.1em}{\end{enumerate} }
\begin{document}
\papertitlepage
\vspace*{0cm}
\Komabanumber{01-03}{hep-th/0107066}{July, 2001}
\Title{A Theorem on the Power of Supersymmetry in Matrix Theory} 
\vspace{1cm}
\Authors{{\sc Y.~Kazama\footnote[2]{kazama@hep3.c.u-tokyo.ac.jp} 
 and T.~Muramatsu
\footnote[3]{tetsu@hep1.c.u-tokyo.ac.jp}
\\ }
\vskip 3ex
 Institute of Physics, University of Tokyo, \\
 Komaba, Meguro-ku, Tokyo 153-8902 Japan \\
  }
\baselineskip .7cm
\Abstract
For the so-called  source-probe configuration in Matrix theory, 
we prove  the following theorem 
 concerning  the power of supersymmetry (SUSY):\  Let $\delta$ be 
a quantum-corrected effective SUSY transformation operator expandable in 
 powers of the coupling constant $g$ as 
$\delta = \sum_{n\ge 0} g^{2n} \delta^{(n)}$, where $\delta^{(0)}$ 
 is of the tree-level form. 
Then, apart from an overall constant, 
the SUSY Ward identity $\delta \Gamma=0$ determines
 the off-shell effective action $\Gamma$ uniquely to 
 arbitrary order of perturbation theory, 
provided that the $SO(9)$ symmetry is preserved. 
Our proof depends only on the properties of the tree-level SUSY 
transformation laws and does not require the detailed knowledge of 
 quantum corrections. 
\newpage
\baselineskip 3.5ex
\section{Introduction}  
 \sectionnumbering
\renewcommand{\thefootnote}{\arabic{footnote}}
Ordinarily, the role of a  global symmetry with a finite number of 
 parameters is to a large extent  kinematical.  It 
provides  conservation laws for the currents and charges, 
relates the Green's functions, restricts the possible form 
 of counter terms and so on, but does not control the dynamics
 completely. The associated Ward identity
 is of an integrated form unlike the more powerful 
  Schwinger-Dyson equation and 
from group-theoretic point of view it only organizes a set of terms 
 into invariants for which independent coupling constants can be 
 assigned. Especially, when one deals with the effective action, which 
 can contain arbitrary number of derivatives of fields, there may be
 many such invariants and  global symmetries are not expected 
 to be able to fix their relative strengths. 
\parsmallskip
Despite these well-known facts, certain theories with a large number of
 global supersymmetries (SUSY) might prove to be exceptions under some
 favorable conditions. Notably, for the so-called Matrix theory 
 for M-theory, proposed by Banks, Fischler, Shenker and Susskind \cite{bfss,
susskind}, 
 evidence has been accumulating \cite{Pabanetal1,Pabanetal2,lowe,ss,
Hyunetal,np,Kaz-Mura1,Kaz-Mura3} that its high degree of 
 SUSY, namely with the maximally allowed 16 supercharges, may be
 powerful enough
 to determine the effective action of a D-particle in interaction with 
 an aggregate of coincident source D-particles, \ie for the so-called
 source-probe configuration. 
Judged from our common wisdom this may sound rather surprising 
 but on the other hand impressive agreement \cite{BBPT, 
Okawa-Yoneya,Okawa-Yoneya2,TR} with 11 dimensional 
 supergravity lends support to such a conjecture, as the latter 
is considered to be unique. 
\parsmallskip
A clear-cut settlement of this issue is hampered by the lack 
 of unconstrained superfield formalism for this system. One is forced 
 to deal with the component formalism, where SUSY is intertwined
 with gauge (BRST) symmetry and  its algebra does not close without 
 the use of the full equations of motion. A related difficulty is that 
 the SUSY transformation laws for the effective action receive complicated
 quantum corrections and one cannot
 easily identify possible SUSY invariants. 
\parsmallskip
In order to gain insight into the issue of the power of 
 SUSY in Matrix theory, in a series of papers \cite{Kaz-Mura1,
Kaz-Mura2,Kaz-Mura3}
we have (i) derived the SUSY Ward identity for the background gauge 
in the form where the quantum-corrected SUSY transformation laws can be 
read off in  closed forms, (ii) computed the effective action $\Ga$ 
and the SUSY transformation operator $\delta$ 
 explicitly for arbitrary off-shell
 trajectory of the probe D-particle including all the 
 spin effects at 1-loop at order 4 in the derivative expansion\footnote{
As usual, the concept of \lq\lq order" in the derivative expansion is
 defined as the number of derivatives plus half the number of 
 fermions.}, 
 (iii) checked that they indeed satisfy the SUSY Ward identity $\delta \Ga=0$,
 and (iv) finally demonstrated that, given such $\delta$, the solution to
 the Ward identity regarded as a functional differential
 equation for $\Ga$ is unique to the order specified above. 
All of these calculations were extremely cumbersome and elaborate codes had
 to be developed for the algebraic manipulation program Mathematica, 
 including a new fast algorithm for generating $SO(9)$ Fierz identities
 of considerable complexity. 
Although  restricted to the source-probe 
situation, the fact that the fully off-shell effective action 
 was uniquely determined from the knowledge of the SUSY transformation 
 laws was remarkable and strongly suggested that 
 this feature would persist to higher orders. 
\parsmallskip
Indeed our result was not an accident that occurred at a particular low order. 
In this article, we shall be able to prove the following theorem valid
 for the source-probe configuration:
\parbigskipn {\it Theorem :} \quad 
Let $\delta$ be 
a quantum-corrected effective SUSY transformation operator expandable in 
 powers of the coupling constant $g$ as 
$\delta = \sum_{n\ge 0} g^{2n} \delta^{(n)}$, where $\delta^{(0)}$ 
 is of the tree-level form. 
Then, apart from an overall constant, 
the SUSY Ward identity $\delta \Ga=0$ determines
 the off-shell effective action $\Ga$ uniquely to 
 arbitrary order of perturbation theory, 
provided that the $SO(9)$ symmetry is preserved. 
\parmedskipn
Actually the main effort will be  devoted to
 proving the following proposition, from which our theorem follows
straightforwardly:
\parbigskipn {\it Proposition :} \quad Let $\delta^{(0)}$ be 
 the tree level SUSY transformation operator. Then assuming that $SO(9)$ 
 is a good symmetry, the solution to the functional differential 
 equation  $\delta^{(0)}\Ga=0$ is unique up to an overall constant, 
 and is given by the tree level action. 
\parsmallskip
We wish to emphasize that our proof of the Proposition and hence the Theorem 
will depend only on the properties of the tree-level SUSY 
transformation laws and will not require the detailed knowledge of 
 quantum corrections. 
\parsmallskip
To avoid any possible misconception, however,  let us state clearly that 
our theorem {\it does not yet} prove that supersymmetry (together with 
 $SO(9)$ symmetry) completely determines the source-probe dynamics 
in Matrix theory. What it states is that given an appropriate $\delta$, 
 which for example can be computed independently of $\Ga$ by 
 the method developed in \cite{Kaz-Mura1} in some gauge, the Ward identity
 fixes  $\Ga$ completely.  In order to claim that SUSY fully determines 
 the dynamics, one must show that it is capable of determining 
 {\it both} $\delta$ {\it and} $\Ga$ simultaneously in a self-consistent
 manner, up to field redefinitions,
 without assuming the knowledge of the underlying action of 
 the Matrix theory. How this more ambitious program should be formulated 
and attempted will be discussed in the final section. 
\parsmallskip
Nevertheless, we believe that our theorem discloses another aspect of 
remarkably powerful features of maximal supersymmetry: 
It clears a highly non-trivial necessary requirement for 
SUSY to be able to \lq\lq determine everything", under certain conditions. 
\parsmallskip
The rest of the article is organized as follows: In Section 2, after 
 setting up our problem, we derive what will be called the \lq\lq basic 
 equation" on which the proof of our Proposition will be based. 
It is a reformulation of the consistency or \lq\lq integrability" condition 
 for the tree-type Ward identity, namely 
 $(\delta^{(0)}_\lam \delta^{(0)}_\ep -\delta^{(0)}_\ep
\delta^{(0)}_\lam)\Ga=0$. Although it provides only a necessary condition, 
 it will prove to be sufficiently restrictive.  Proofs of the Proposition 
 and the Theorem will be given in Section 3. As the proofs are somewhat 
 involved, we will first illustrate our basic ideas using a simple example. 
 These ideas are then sharpened and extended into a number of 
 Lemmas which apply for general situation. These Lemmas step by step 
 reduce the possible solutions to be of more and more restricted form and 
 lead  to the proof of the Proposition.  This in turn 
 immediately yields our main Theorem, by a recursive argument
 with respect to the  powers of the coupling constant. Finally in Section 4, 
we will discuss 
 and indicate the line of attack for the important remaining problems, 
namely the investigation of the multi-body case and of the possibility 
 of  complete determination of the dynamics by SUSY. 
\section{Derivation of the Basic Equation}  
 \sectionnumbering
\subsection{Preliminaries}
Let us begin by briefly recalling the relevant features of Matrix theory
 and set up the problem. 
\parsmallskip
The classical action for the $U(N+1)$
 Matrix theory in the Euclidean formulation 
 is given by 
\begin{align}
S_0 &= \trace \int\!\!d\tau  \ \Biggl\{ \frac{1}{2} \com{D_\tau}{X_m} ^2 
- \frac{g^2}{4}
 \com{X_m}{X_n} ^2  \nn\\
& \qquad +\frac{1}{2}\Theta^T \com{D_\tau}{\Theta} 
-\frac{1}{2} g \Theta^T \ga^m\com{X_m}{\Theta} \Biggr\}\comma \\
D_\tau &\equiv \del_\tau -igA \period
\end{align}
In this expression, 
$X^m_{ij}(\tau), A_{ij}(\tau)$ and $\Theta_{\al,ij}(\tau)$ are 
the $(N+1)\times (N+1)$ 
hermitian matrix fields, representing the bosonic part of the
D-particles, the gauge fields, and the fermionic part of the D-particles,
 respectively.  $\ga^m$ are the real symmetric $16\times 16$ $SO(9)$ $\ga$-matrices, and 
the vector index $m$ runs from $1$ to $9$. 
\parsmallskip
This action is known to possess a number of important symmetries. 
Among them, the main focus in this article will be the supersymmetry, 
 which carries 16 spinorial parameters 
$\ep_\al$ and transforms  the basic fields in the manner
\begin{align}
\delta_\ep A &= \ep^T \Theta \comma \qquad 
\delta_\ep X^m = -i\ep^T \ga^m\Theta \comma \label{treesusy}\\
\delta_\ep \Theta &= i\left( \com{D_\tau}{X_m} \ga^m
 +{g\over 2} \com{X_m}{X_n} \ga^{mn} \right) \ep \period
\end{align}
Although the algebra closes only on-shell up to field-dependent 
 gauge transformations, $S_0$ itself is invariant without the use of 
equations of motion, \ie off-shell. 

We will also make use of the 
 $SO(9)$ symmetry throughout for restricting the forms of possible terms 
 to be considered. Its  role, however,  is relatively minor 
as we will not need to resort to a host of Fierz
 identities of considerable complexity, which were heavily utilized in 
 our previous works. 
\parsmallskip
In this article, we shall concentrate on the so-called source-probe 
situation, namely the configuration of a probe D-particle 
interacting with a large number, $N$, of the source D-particles all sitting 
 at the origin. This is expressed by the splitting 
\begin{alignat}{2}
X_m(\tau) &= {1\over g} B_m(\tau) +Y_m(\tau) \comma &  
\qquad \Theta_\al(\tau) &= {1\over g} \theta_\al(\tau)
+ \Psi_\al(\tau)\comma  \\
B_m(\tau) &= {\rm diag}\, (r_m(\tau), \overbrace{0,0,\ldots , 0}^N) \comma &
\qquad \theta_\al(\tau) &=  {\rm diag}\, 
 (\theta_\al(\tau), \overbrace{0,0,\ldots , 0}^N) \comma 
\end{alignat}
where $B_m(\tau)$ and $\theta_\al(\tau)$ are 
 the bosonic and the fermionic backgrounds expressing the 
positions and the spin degrees of freedom of the D-particles 
 respectively 
 and $Y_m(\tau)$ and $\Psi_\al(\tau)$ denote the quantum 
fluctuations around them. 
We will be interested in the general case
 where $r_m(\tau)$ and $\theta_\al(\tau)$ are {\it arbitrary functions}
 of $\tau$ not satisfying equations of motion. 
\parsmallskip
At the tree level, the action and the SUSY transformations for 
 the probe D-particle are  of the form
\begin{eqnarray}
\Gazero
 &=& \int d\tau \left( {\dot{r}^2\over 2g^2} + {\theta\thdot \over 2g^2}\right)
\comma \label{gaz}\\
\dzep  r_m &=& {1\over i}\ep \ga_m \theta \comma \label{dzr}\\
\dzep \theta_\al &=& i (\dot{\rslash} \ep)_\al \comma \label{dzth}
\end{eqnarray}
where the dot means differentiation with respect to $\tau$ and 
 $\theta\thdot$ stands for $\theta_\al
\thdot_\al$, etc. 
After a gauge is fixed and the functional integral over the 
 fluctuations is performed, one obtains the quantum effective action 
\begin{eqnarray}
\Ga 
 = \int d\tau \calL \left[r_m, \theta_\al\right]  \period
\end{eqnarray}
Although a general argument does not exist, explicit calculations
 performed so far in the so-called  \lq\lq background gauge" strongly 
 indicate that $\calL\left[r_m,\theta_\al\right]$ is a sum 
 of  local products of derivatives of $r_m(\tau)$ and $\theta_\al(\tau)$.
As it will be emphasized again at the beginning of Sec.~3, our proof of 
 the Theorem and the Proposition will operate at each order in $g^2$ 
{\it and} in the derivative expansion, so that effectively 
we will be dealing with a finite number of local expressions at each step 
regardless of whether the entire $\calL$ is fully local or not. 

$\Ga$ satisfies the SUSY Ward identity 
\begin{eqnarray}
\delta_\ep \Ga =0 \comma \label{wardid}
\end{eqnarray}
where $\delta_\ep$ is the quantum-corrected effective SUSY transformation 
operator. 
In a previous article \cite{Kaz-Mura1}, we have derived the explicit
 form of this Ward identity in the background gauge, where $\delta_\ep$
 is given in closed form in terms of expectation values of 
 certain products of fields.  Further in \cite{Kaz-Mura2} a 
complete calculation of $\Ga$ and $\delta_\ep$ at 1-loop at order 4 
 in the derivative expansion was performed and the validity of 
 (\ref{wardid}) was explicitly verified at that order. 
\parsmallskip
As was stated in the introduction, an interesting question is, 
for a given $\delta_\ep$ how unique  $\Ga$ is,
 without the use of the equations 
 of motion. For the simplest case, namely at  1-loop at order 2,
 it was found that $\Ga$ is completely determined by 
$\delta_\ep$ \cite{Kaz-Mura1}. Moreover, a highly non-trivial calculation
 at order 4  again showed that $\Ga$ is unique \cite{Kaz-Mura3}. In fact, 
these results were not accidental:
 We shall prove that this feature persists to an arbitrary order of
 perturbation theory. 
\subsection{Consistency equation}
Since our main Theorem can be shown to follow
 rather easily from  the Proposition, we will first concentrate on 
 the proof of the latter. 
\parsmallskip
What we wish to show is that the solution to the equation 
$\dzep \Ga=0$ is unique up to an overall constant, where the action of
 $\dzep$ is as defined in (\ref{dzr}) and (\ref{dzth}). 
This equation, despite its simple appearance, is a highly 
complicated partial differential equation involving Grassmann variables
 for the integrand $\calL$ of $\Ga$.  In fact, written out in full, 
 it reads
\begin{eqnarray}
\dzep \Ga &=& \int d\tau \left(  \dzep r_m(\tau) 
{\delta  \Ga \over \delta r_m(\tau)} +
\dzep \theta_\al(\tau) {\delta\Ga  \over \delta \theta_\al(\tau)}\right)
=0 \comma \label{dzepga}
\end{eqnarray}
where, in terms of $\calL$,  the functional derivative $\delta \Ga/\delta r_m$ 
 means (suppressing $\tau$-dependence) 
\begin{eqnarray}
{\delta \Ga \over \delta r_m} &=& 
\calD[\calL, r_m] \comma \\
\calD[\calL, r_m]  &\equiv& {\del \calL \over \del
 r_m} -\del_\tau {\del \calL \over \del \dot{r}_m} + 
 \del_\tau^2 {\del \calL \over \del \ddot{r}_m} + \cdots \comma
\label{functint}
\end{eqnarray}
and similarly for $\delta \Ga/\delta\theta_\al$. Later we will often 
use the symbol $\calD_N[\calL, r_m]$
 to denote $\calD[\calL,r_m]$ truncated at
 (and including) $(-1)^N\del_\tau^N (\del \calL/\del  r^{(N)}_m)$,
 where $r^{(N)}_m$ is the $N$-th derivative of $r_m$ with respect to
 $\tau$. 
\parsmallskip
Our basic idea 
 for analyzing (\ref{dzepga}) is to look  at its consistency or 
 \lq\lq integrability" equation, namely 
\begin{eqnarray}
(\dzlam\dzep -\dzep\dzlam) \Ga =0 \period\label{intcond}
\end{eqnarray}
Although this only gives a necessary condition, it will prove to be 
 powerful enough to fix the fermionic part 
of $\Ga$ completely.  The original equation $\delta^{(0)}_\ep \Ga=0$
 then immediately determines the bosonic part as well. 
\parsmallskip
Now computing the left hand side of (\ref{intcond}) by using the 
 functional derivative as in (\ref{dzepga}), one finds that 
 the parts containing the second functional derivatives of $\Ga$ 
 cancel and the result is 
\begin{eqnarray}
(\dzlam \dzep -\dzep \dzlam) \Ga &=&
 \int d\tau \left( [\dzlam, \dzep]r_m(\tau) {\delta \Ga \over 
\delta r_m(\tau)} + [\dzlam, \dzep]\theta_\al(\tau) {\delta \Ga \over 
\delta \theta_\al(\tau)} \right) \period \label{intcond2} \nn\\
\end{eqnarray}
The closure of the algebra $[\dzlam, \dzep]$ is easily calculated. 
On $r_m$ it simply gives the usual time-translation 
\begin{eqnarray}
[\dzlam, \dzep] r_m &=& 2(\ep\lam) {d r_m \over d\tau} \comma 
\end{eqnarray}
while on $\theta_\al$ one finds 
\begin{eqnarray}
[\dzlam, \dzep] \theta_\al &=& (\lam \ga_m \thdot)(\ga_m \ep)_\al 
 -(\ep \ga_m \thdot)(\ga_m \lam)_\al \period
\end{eqnarray}
By using a well-known  $SO(9)$ $\ga$-matrix identity
\begin{eqnarray}
\ga^m_{\al\be}\ga^m_{\ga\delta} + (\mbox{cyclic in 
 $\al,\be,\ga$}) =\delta_{\al\be} \delta_{\ga\delta} +
(\mbox{cyclic in 
 $\al,\be,\ga$}) \comma
\end{eqnarray}
it can be rewritten as
\begin{eqnarray}
[\dzlam, \dzep] \theta_\al &=&2(\ep\lam){d \over d\tau} \theta_\al 
 + \Theta_\al(\lam,\ep) \comma 
\end{eqnarray}
where
\begin{eqnarray}
\Theta_\al(\lam,\ep) &\equiv & (\lam \ga^m \ep)
(\ga_m \thdot)_\al + \ep_\al(\lam \thdot) -\lam_\al (\ep\thdot)
 +(\lam\ep) \thdot_\al \period
\end{eqnarray}
Let us put these results back into (\ref{intcond2}). Then, by using 
 the definition of the functional derivative (\ref{functint}) and 
 integration by parts, one finds that the contributions from 
 the canonical time-translation part of the closure relation add up 
to an integral of a total derivative as 
\begin{eqnarray}
2(\ep\lam) \int d\tau \left( {dr_m(\tau) \over d\tau}
 {\delta \Ga \over \delta r_m(\tau)}
 + {d\theta_\al(\tau) \over d\tau}
 {\delta \Ga \over \delta \theta_\al(\tau)}\right)
&=& 2(\ep\lam) \int d\tau {d \calL \over d \tau} 
\end{eqnarray}
and vanish. Thus, the consistency equation is reduced to 
\begin{eqnarray}
\int d\tau \Theta_\al(\lam, \ep) F_\al=0 \label{intcond3} \comma 
\end{eqnarray}
where for simplicity we defined 
$F_\al \equiv \delta \Ga /\delta \theta_\al$. This notation will be 
 used throughout the rest of the article. 
Further,  since this equation must hold for arbitrary $\lam$ and $\ep$,
 we may remove  these global spinor parameters and obtain
\begin{eqnarray}
\int d\tau\, \calF_{\be\ga} &=& 0 \comma \\
\calF_{\be\ga} &\equiv & -\ga^m_{\be\ga} (\thdot \ga_m F) + \thdot_\ga
F_\be + \thdot_\be F_\ga -\delta_{\be\ga}(\thdot F)  \period
\label{intcond4}
\end{eqnarray}
There are two ways that this equation may be satisfied; either 
(a)\ $\calF_{\be\ga}$ vanishes identically or (b)\ $\calF_{\be\ga}$ is 
 a total derivative. 
\parsmallskip
Let us begin with case (a). First, take the trace over the index set 
$(\be,\ga)$. Since $\ga^m$ is traceless, we immediately get 
$\thdot F=0$. Putting this back into the equation, we have 
$-\ga^m_{\be\ga} (\thdot \ga_m F) + \thdot_\ga
F_\be + \thdot_\be F_\ga  =0$. Now contract this with $\ga^n_{\ga\be}$. Then, using $\Tr \ga^n\ga^m=
16 \delta^{mn}$, we find $\thdot \ga_n F=0$. In this way, the consistency
 equation is reduced to the following simple form:
\begin{eqnarray}
\thdot_\ga F_\be + \thdot_\be F_\ga =0 \period \label{basiceq1}
\end{eqnarray}
Note that this guarantees the previous two relations, 
$\thdot F=\thdot \ga_n F=0$, as well. 
\parsmallskip
There are two types of possible solutions to this algebraic equation. 
One is of the form 
\begin{eqnarray}
F_\al = \thdot_\al G\comma \label{fg}
\end{eqnarray}
where $G$ is some $SO(9)$ scalar. Due to the Grassmann nature of 
$\thdot$, this obviously satisfies (\ref{basiceq1}). The other 
 is the case where $F_\ga$ contains the product of 16 $\thdot_\sigma$'s, 
 namely 
\begin{eqnarray}
\Xi \equiv \prod_{\sigma=1}^{16}\thdot_\sigma
 \comma \label{Xi}
\end{eqnarray}
which is easily checked to be $SO(9)$ invariant. In this case, 
 $\thdot_\be F_\ga$ vanishes identically for any $\be$ and $\ga$ and hence
 (\ref{basiceq1}) is trivially satisfied. 
\parsmallskip
Now let us turn to case (b), for which $\calF_{\be\ga} = dG_{\be\ga}/d\tau$ 
 for some non-constant $G_{\be\ga}$. Going through exactly 
 the same procedure as in case (a), we find 
\begin{eqnarray}
\thdot_\ga F_\be + \thdot_\be F_\ga 
&=& {d  \tilde{G}_{\be\ga} \over d\tau}\comma \\
\tilde{G}_{\be\ga} &=& G_{\be\ga} -{1\over 14} \delta_{\be\ga}
\Tr G -{1\over 14} \ga^m_{\be\ga} \Tr(\ga_m G) \period
\end{eqnarray}
Rewriting $\thdot_\ga F_\be + \thdot_\be F_\ga =
{d \over d\tau}\left( \theta_\ga F_\be +\theta_\be F_\ga\right)
 -(\theta_\ga \dot{F}_\be +\theta_\be \dot{F}_\ga)$, we must demand that 
the expression 
$\theta_\ga \dot{F}_\be +\theta_\be \dot{F}_\ga$ either (b-1)\  vanish identically or (b-2)\ be a total derivative itself. 

Let us quickly dispose of the possibility (b-1). Since $\dot{F}_\be$, 
 being a time derivative, cannot be proportional to the 
 product $\prod_{\sigma=1}^{16} \theta_\sigma$,  the case (b-1) 
 can occur only if $\dot{F}_\be =\theta_\be Z$ for some $Z$. But clearly 
 it is impossible to produce the structure like $\theta_\be Z$ 
  by a time derivative of a local\footnote{Remember that implicitly 
 we are dealing with expressions  \lq\lq order by order" in the double sense
and hence locality argument is valid.} expression. Hence case (b-1) is 
 excluded. 

Now consider case (b-2). 
For this to be possible at all, $F_\ga$ must contain the derivative 
 of $\theta_\ga$, namely it must be of the form 
$F_\ga=\thdot_\ga f$ for some $SO(9)$ invariant $f$. 
Then, we have $\theta_\ga \dot{F}_\be +\theta_\be \dot{F}_\ga
 = (\theta_\ga\thdot_\be+\theta_\be \thdot_\ga)f$. As this does not
 contain $\dot{f}$, needed to form a total derivative, 
$f$ can only be a constant. This reduces the possibility to the 
 expression 
$\theta_\ga\thdot_\be+\theta_\be \thdot_\ga$. But this is clearly not 
 a total derivative and hence case (b-2) is also excluded. 
\parsmallskip
Summarizing, we have shown that only case (a) can be possible and 
 the consistency equation is reduced to 
\begin{eqnarray}
{\delta \Ga \over \delta \theta_\al}
 &=& \calD[\calL,\theta_\al] = \thdot_\al G + \Xi H_\al \comma 
\label{basiceq}
\end{eqnarray}
where $\Xi$ is as defined in (\ref{Xi}) and 
$G$ and $H_\al$ are, respectively, an $SO(9)$ scalar and 
 a spinor.
 Note that compared to the original 
 equation $\dzep \Ga=0$, which contains functional derivatives 
  with respect to both $r_m$ and $\theta_\al$, the consistency equation
above only involves the one with respect to the latter and hence 
 is more tractable.
 Hereafter we shall call (\ref{basiceq}) our {\it basic equation}. 
In the next section, we shall solve this basic equation completely 
 to prove our Proposition, which will then straightforwardly lead to 
 our main Theorem. 
\section{Proof of the Theorem}  
 \sectionnumbering
{}From the preceding analysis, we have learned that as a necessary 
 condition a solution $\Ga$ to the tree-type Ward identity $\dzep \Ga=0$
 must satisfy the basic equation (\ref{basiceq}). Below we shall 
 prove that, up to an overall constant and $\theta$-independent
 structures, the solution to (\ref{basiceq})
 is unique and is given by $\int d\tau \theta\thdot$, which is nothing 
 but the fermionic part of the tree level action $\Ga^{(0)}$. Then imposition
 of $\dzep \Ga=0$ immediately fixes 
the $\theta$-independent part to coincide with 
 the bosonic part of $\Ga^{(0)}$.
\parsmallskip
Since the 
 proof is rather involved, we shall first illustrate our basic 
 strategy by treating a simple example. Then, we shall develop 
 several useful lemmas, which are basically the generalizations 
 of the logic used for that example. Finally with the aid of these lemmas
 the proof of the Proposition and the Theorem will be given. 
\parsmallskip
Before we begin our proof, let us make some important remarks. 
(i)\ Since we shall work to an arbitrary
 but finite order in $g^2$ and at each such order the space of solutions
 can be further classified by the order in the sense of derivative expansion,
  we may assume, without loss of generality, 
 that $\calL$ consists of a {\it finite number of local expressions.} 
In particular, this implies 
 that the number of $\tau$-derivatives acting on $\theta_\al$ has a maximum
 in $\calL$. Hereafter we shall refer to such a spinor with the largest 
number of derivatives as \lq\lq SPLD". 
(ii)\ As the actual proof does not depend on which order of 
$g^2$ one is working, 
we shall not display $g^2$-dependence. (iii)\ Lastly, $SO(9)$ invariance 
will be  taken for granted throughout. 
\subsection{Illustration of a Simple Case}
Consider the simple case where $\calL$ consists of 
 bilinears of $\theta$ and its derivatives up to $\thddot$ only. 
The general form of such $\calL$ is 
\begin{eqnarray}
\calL &=& \half \theta A \theta + \theta B \thdot + \half \thdot C \thdot
+\theta D \thddot + \thdot E \thddot + \half \thddot F \thddot \comma
\end{eqnarray}
where $A\sim F$ are bispinors made out of $r_m$, its derivatives,
  $SO(9)$ $\ga$-matrices and the unit matrix. 
Clearly, $A,C$ and $F$ must be antisymmetric.  In this case, the 
 special product $\Xi$ cannot occur and the equation 
 to be solved is of the form 
\begin{eqnarray}
\calD_2[\calL,\theta_\al]=\thdot_\al G \label{exbeq}\comma 
\end{eqnarray}
where $\calD_2[\calL, \theta_\al]$ is given by  
$\del \calL/\del \theta_\al 
-\del_\tau (\del \calL / \del 
 \thdot_\al)+\del_\tau^2 (\del \calL/ \del \thddot_\al)$.
 We now proceed in steps. 
\parmedskipn
1.\quad  First, focus on 
 the \lq\lq $F$" term, containing more than one 
SPLD ($\thddot$ in this example). In $\calD_2[\calL;\theta_\al]$, 
 look at the term containing $\thd{4}$. 
Such a term can only be produced from the \lq\lq $F$" term through 
$\del_\tau^2 (\del \calL/\del \thddot_\al)$ and 
it cannot be canceled by anything
 else. Hence $F$ must vanish. 
\parsmallskipn
2.\quad Next, consider the term which is 
linear in $\thddot$ and contains 
 a spinor with one less derivative, namely $\thdot$. In this example it is 
 the \lq\lq $E$" term. Look at the terms containing $\thd{3}$ in 
 $\calD_2[\calL;\theta_\al]$. 
They come from $-\del_\tau (\del \calL /\del \thdot_\al)$
 and from $\del_\tau^2 (\del \calL /\del \thddot_\al)$ and add up to give 
$-(E+E^T)\thd{3}$. As this must vanish, $E$ must be 
antisymmetric. Then, the \lq\lq $E$" term can be rewritten as
\begin{eqnarray}
\thdot E \thddot &=& \del_\tau(\thdot E \thdot) +\thdot E^T \thddot 
 -\thdot \dot{E} \thdot \period
\end{eqnarray}
On the right-hand side (RHS),
 the first term is a total derivative and can be dropped, 
 while  due to antisymmetry the second term is actually the negative
of the left-hand side (LHS). 
Thus we get
\begin{eqnarray}
\thdot E \thddot &=& -\half \thdot \dot{E} \thdot \period
\end{eqnarray}
This does not contain $\thddot$ any more and can be absorbed in the 
 \lq\lq $C$" term. 
\parsmallskipn
3.\quad  Having disposed of the \lq\lq $E$" term, look at the 
 remaining term with $\thddot$, namely the \lq\lq $D$" term. It 
 can be rewritten as
\begin{eqnarray}
\theta D \thddot &=& \del_\tau(\theta D \thdot) -\thdot D\thdot -\theta
 \dot{D} \thdot \period
\end{eqnarray}
Thus, $\thddot$ arises only from the total derivative term, which we 
 may discard, and the \lq\lq $D$" term can be absorbed by the other terms 
 with less number of derivatives on the spinors. At this stage
 all the terms containing SPLD have disappeared. 
\parsmallskipn
4.\quad  Next consider the \lq\lq $C$" term, with two powers of 
$\thdot$, and look for terms with $\thddot$ in $\calD_2[\calL;\theta_\al]$. 
It is produced only from $-\del_\tau (\del \calL /\del \thdot_\al)$ and is 
 given by $-C\thddot$. This must vanish and hence $C=0$. 
\parsmallskipn
5.\quad  We are now left with $\calL = (1/2)\theta A\theta +\theta B \thdot$. 
The \lq\lq $B$" term can be rewritten as
\begin{eqnarray}
\theta B \thdot &=& \del_\tau (\theta B \theta) +\theta B^T \thdot -\theta 
 \dot{B}\theta \period
\end{eqnarray}
Contributions from the symmetric part of $B$, to be denoted by $B_S$, 
 cancel between LHS and RHS. Writing the antisymmetric part as $B_A$, 
we get
\begin{eqnarray}
\theta B_A \thdot &=& \half \left( \del_\tau (\theta B_A \theta)
-\theta  \dot{B_A}\theta \right) \period
\end{eqnarray}
Thus,  dropping the total derivative, this can be absorbed 
by the \lq\lq $A$" term. 
So we only need to keep $B_S$,  and $\calL$ is reduced to
$(1/2)\theta A \theta + \theta B_S \thdot$. 
\parsmallskipn
6.\quad  Finally, look at the basic equation (\ref{exbeq}). 
It becomes 
\begin{eqnarray}
(A\theta)_\al +2(B_S \thdot)_\al +(\dot{B}_S\theta)_\al =\thdot_\al G
\period
\end{eqnarray}
Since $A$ is antisymmetric whereas $B_S$ is symmetric, the terms 
 proportional to $\theta$ must vanish separately. Thus, we find $A=0$ and 
 $B_S=$constant. Now, since the only symmetric constant bispinor is 
$\delta_{\al\be}$, we must have ${B_S}_{\al\be} = c\delta_{\al\be}$, with 
 $c$  a constant. This in turn dictates $G=2c$. 
In this way, we find the unique solution 
\begin{eqnarray}
\calL &=& c (\theta\thdot) \period
\end{eqnarray}
This completes the proof for this simple case. 
\subsection{Useful Lemmas}
In this subsection, we shall prove a number of useful lemmas to pave the way 
 for the proof of the Proposition. The proofs of these lemmas may 
 look rather involved but essentially
 they are refined generalization of the logic used for the 
 simple case described above. 
\parsmallskip
The first lemma eliminates the possibility of the appearance of 
 the special product $\Xi$ under certain conditions. 
\Lemma{1} Let $\thd{N}$ be the  SPLD in $\calL$. Then
 for $N \ge 2$, the special product $\Xi$ cannot appear in $\calL$. 
\Proof 
The product $\Xi$ can be written as
\begin{eqnarray}
\Xi &=& \thdot_{\al_1}\thdot_{\al_2}\cdots \thdot_{\al_{16}} E_{\al_1\al_2
\ldots \al_{16}} \comma \\
 E_{\al_1\al_2\ldots \al_{16}} &\equiv& {1\over 16!}\ep_{\al_1\al_2\ldots
 \al_{16}}\period
\end{eqnarray}
Then, if $\calL$ contains $\Xi$, it must be of the form  
\begin{eqnarray}
\calL &=& \Xi\, \thd{N}_{\be_1}\thd{N}_{\be_2}\cdots \thd{N}_{\be_n}
Z_{\be_1\be_2\ldots \be_n} \nn\\
&=& \thdot_{\al_1}\thdot_{\al_2}\cdots \thdot_{\al_{16}}
\thd{N}_{\be_1}\thd{N}_{\be_2}\cdots \thd{N}_{\be_n}
 E_{\al_1\al_2\ldots \al_{16}}Z_{\be_1\be_2\ldots \be_n}\comma 
\end{eqnarray}
where $Z_{\be_1\be_2\ldots \be_n}$ may contain derivatives of the spinor up to $\thd{N-1}$ except 
 for $\thdot$'s, which are explicitly exhibited. 
Now look at the terms of the form $(\thdot)^{15} \thd{N+1} (\thd{N})^{n-1}$
 in $\calD_N[\calL,\theta_\al]$. 
There are only two sources for such a structure. One is produced as
\begin{eqnarray}
-\del_\tau {\delta \calL \over \delta \thdot_\al} &\ni&
 -16n \thdot_{\al_2} \cdots
 \thdot_{\al_{16}}
 \thd{N+1}_{\be_1} \thd{N}_{\be_2} \cdots \thd{N}_{\be_n}
 E_{\al \al_2 \ldots \al_{16}} Z_{\be_1
\be_2\ldots \be_n} \period \label{lem1eq1}
\end{eqnarray}
The other comes from
\begin{eqnarray}
(-1)^N \del_\tau^N {\delta \calL\over \delta \thd{N}_\al}
 &\ni & (-1)^{N+16} 16n \thdot_{\al_1}\cdots \thdot_{\al_{15}}
\thd{N+1}_{\al_{16}} 
\thd{N}_{\be_2}\cdots \thd{N}_{\be_n} 
E_{\al_1\ldots \al_{16}}Z_{\al \be_2 \ldots \be_n}\period \label{lem1eq2}
\nn\\
\end{eqnarray}
Since the product $E_{\al_1\al_2\ldots \al_{16}}Z_{\be_1\be_2\ldots \be_n}$
cannot be totally antisymmetric in its entire index set 
 $(\al_1, \ldots, \al_{16},\be_1\ldots,\be_n)$, (\ref{lem1eq1})
 and (\ref{lem1eq2}) are independent and cannot
 cancel each other.
 Moreover, neither term can be proportional to $\thdot_\al$ due to 
 the total antisymmetry of $E_{\al_1\ldots \al_{16}}$. (For (\ref{lem1eq1})
 it is obvious. For (\ref{lem1eq2}), if there exists
 $\delta_{\al \al_1}$, then 
 there must also be $\delta_{\al\al_{16}}$, which produces 
 $\thd{N+1}_\al$ instead of $\thdot_\al$.)
Therefore they must vanish separately  and we find  $Z_{\be_1
\ldots \be_n}=0$, which proves the lemma. \qed
\parsmallskip
The second lemma will severely restrict the way a product of 
more than two SPLD's can appear in the effective Lagrangian $\calL$.
\Lemma{2} 
Let  $\thd{N}$, with $N\ge 2$,  be the SPLD in $\calL$ and assume 
 that there are at least two of them. 
Then, they can only be contained through the special combination
 $\thdot\thd{N}$. 
\Proof In general such structures with different number of $\thd{N}$'s 
may appear in $\calL$. 
However, as they lead to algebraically independent
 expressions in the basic equation,
 we may treat each term separately. Thus, we consider 
 a generic structure of that type and write it as  
\begin{eqnarray}
\calL &=& \thdot_{\be_1} \thdot_{\be_2} \cdots \thdot_{\be_m}
 \thd{N}_{\ga_1} \thd{N}_{\ga_2} \cdots\thd{N}_{\ga_n} 
Z_{\be_1\be_2 \ldots \be_m;\ga_1\ga_2 \ldots \ga_n} \comma 
\end{eqnarray}
where $Z_{\be_1\be_2 \ldots \be_m;\ga_1\ga_2 \ldots \ga_n}$ does not contain $\thdot$ nor $\thd{N}$ and is separately antisymmetric in the indices 
$\{\be_1\be_2 \ldots \be_m\}$ and in $\{\ga_1\ga_2 \ldots \ga_n\}$. 
Now look at the structure  containing $\thd{2N}$ in 
$\calD_N[\calL,\theta_\al]$.
Such a structure can only be
 produced from $\del_\tau^N (\del \calL/\del \thd{N})$ and  is proportional to
\begin{eqnarray}
\thdot_{\be_1} \thdot_{\be_2} \cdots \thdot_{\be_m}
 \thd{N}_{\ga_1} \thd{N}_{\ga_2} \cdots\thd{N}_{\ga_{n-2}} \thd{2N}_{\ga_{n-1}}
Z_{\be_1\be_2 \ldots \be_m;\ga_1\ga_2 \ldots \ga_{n-1} \al}  \period
\end{eqnarray}
Since the appearance of $\Xi$ is excluded by Lemma 1, 
for this to be allowed  it must be proportional to $\thdot_\al$. This 
 requires the existence of $\delta_{\be_i\al}$ in 
$Z_{\be_1\be_2 \ldots \be_m;\ga_1\ga_2 \ldots \ga_{n-1} \al}$ 
and similar Kronecker 
$\delta$'s required by its antisymmetry property. This means that 
 $m \ge n$ and all the indices $\{\ga_1, \ldots, \ga_{n-1},\al\}$
must be paired off with (a part of)
 $\be_i$'s through such Kronecker $\delta$'s. Hence, 
 $\thd{N}$ in $\calL$ can only exist 
through the combination  $\thdot \thd{N}$. \qed
\parsmallskip
The third lemma will further restrict the possible dependence on 
 SPLD to the linear level.
\Lemma{3} Let $\thd{N}$, with $N\ge 2$,  be the SPLD in $\calL$.
 Then, $\calL$ can depend on $\thd{N}$ only linearly. 
\Proof Assume that $\calL$ has more than two $\thd{N}$'s. 
According to Lemma 2, its dependence on $\thd{N}$ must
 be through the combination $\thdot\thd{N}$ and therefore $\calL$ must be of 
form
\begin{eqnarray}
\calL &=& \sum_{m\ge 2} (\thdot\thd{N})^m 
Z_m \comma
\end{eqnarray}
where $Z_m$ contains derivatives of $\theta$ only up to $\thd{N-1}$. 
We now look at the part involving $\thd{2N-1}$ and $\thddot_\al$ in 
$\calD_N[\calL,\theta_\al]$. We will distinguish two cases. 

First suppose $Z_m$ does not contain a term of the form 
$(\thd{N-1}\thddot)^n$. 
Then, the only way that such a structure is produced is through 
 $\del_\tau^N \del \calL/\del \thd{N}_\al$ and is given by
\begin{eqnarray}
\del_\tau^N \dldal{\thd{N}} &\ni & -Nm(m-1) \thddot_\al (\thdot\thd{2N-1})
(\thdot\thd{N})^{m-2} Z_m \period
\end{eqnarray}
Since this is not proportional to $\thdot_\al$ nor to $\Xi$ (
 whose existence has already been excluded by Lemma 1) and  since $m\ge 2$, 
this must vanish and hence $Z_m=0$. 

On the other hand, if $Z_m$ contains $(\thd{N-1}\thddot)^n$, there will be 
 another source. In such a case, we may write $\calL$ in the form
\begin{eqnarray}
\calL &=& \sum_{m\ge 2, n} (\thdot\thd{N})^m (\thd{N-1}\thddot)^n Z_{mn}
\period
\end{eqnarray}
Suppressing the summation symbol, the contribution of the previous type 
 takes the form
\begin{eqnarray}
\del_\tau^N \dldal{\thd{N}} &\ni & -Nm(m-1) \thddot_\al 
(\thdot\thd{2N-1}) (\thdot\thd{N})^{m-2} (\thd{N-1}\thddot)^n Z_{mn}
\comma 
\end{eqnarray}
while the additional contribution is 
\begin{eqnarray}
-\del_\tau^{N-1} \dldal{\thd{N-1}} &\ni & -m'n'\thddot_\al 
(\thdot\thd{2N-1})(\thdot\thd{N})^{m'-1}(\thd{N-1}\thddot)^{n'-1} Z_{m'n'} 
\period
\end{eqnarray}
These contributions must add up to vanish. Therefore, we must have
\begin{eqnarray}
m' &=& m-1\comma \qquad n'=n+1 \comma \\
Nm(m-1)Z_{mn} &=&-m'n'Z_{m'n'} \period 
\end{eqnarray}
{}From this we get
\begin{eqnarray}
Nm Z_{mn} &=& -(n+1)Z_{m-1,n+1} \period
\end{eqnarray}
Now let the largest $m$ and $n$ be $m_{max}$ and $n_{max}$ respectively.
{}From the above equation, we find that $Z_{m_{max},n_{max}}$ must 
 be proportional to $Z_{m_{max}-1, n_{max}+1}$, which however does not exist. 
Thus  $Z_{m_{max},n_{max}}$ must vanish. This contradicts 
 the assumption that $m_{max}, n_{max}$ are the maximum values for $m,n$. 
Applying this logic recursively, we conclude that
 $Z_{mn}$ must vanish for all $m,n$ and $\calL$ may depend on $\thd{N}$ only 
 linearly. \qed
\parsmallskip
The last lemma will completely dispose of SPLD with more than two 
 time derivatives. 
\Lemma{4} Let the SPLD in $\calL$ be $\thd{N}$
 with $N\ge 2$  and assume
 that  $\calL$ depends on $\thd{N}$  linearly. Then, the dependence of 
 $\calL$ on $\thd{N}$ can only  be through a total derivative. 
\Proof We shall distinguish three cases. \parsmallskipn
{\it Case 1:} \quad First assume that $\calL$ contains more than two
 $\thd{N-1}$. Then it can be written as
\begin{eqnarray}
\calL &=& \sum_{m\ge 2}\thd{N-1}_{\be_1}\thd{N-1}_{\be_2}\cdots
 \thd{N-1}_{\be_m}
\thd{N}_\ga Z_{\be_1\be_2 \ldots \be_m;\ga}\comma 
\end{eqnarray}
where $Z_{\be_1\be_2 \ldots \be_m;\ga}$ contains the derivative of 
 $\theta$ only up to $\thd{N-2}$.
 By \lq\lq integration by parts",
 this can be rewritten as (suppressing the summation symbol $\sum_m$ )
\begin{eqnarray}
\calL &=&  \del_\tau \left( \thd{N-1}_{\be_1}\thd{N-1}_{\be_2}\cdots 
\thd{N-1}_{\be_m}
\thd{N-1}_\ga Z_{\be_1\be_2 \ldots \be_m;\ga}\right) \nn\\
&& -m \thd{N}_{\be_1}\thd{N-1}_{\be_2}\cdots 
\thd{N-1}_{\be_m}
\thd{N-1}_\ga Z_{\be_1\be_2 \ldots \be_m;\ga}+ \tilde{Z}\comma 
\end{eqnarray}
where the SPLD in $\tilde{Z}$ is  $\thd{N-1}$. Now if 
$Z_{\be_1\be_2 \ldots \be_m;\ga}$ is not totally antisymmetric 
 with respect to the indices $(\be_2, \be_3, \ldots, \be_m, \ga)$, then
 the second term vanishes and the lemma is proved. 
If it happens to be totally antisymmetric, we shall write
 it as $Z_{\be_1\be_2 \ldots \be_m \ga}$ without the semicolon and 
 the second term on the RHS is seen to have exactly the same structure 
 as $\calL$. In fact it is precisely $-m\calL$. Solving for $\calL$, we get
\begin{eqnarray}
\calL = {1\over m} (\del_\tau(\ \ast\ ) + \tilde{Z})\comma 
\end{eqnarray}
 which proves the lemma. 
\parmedskipn
{\it Case 2:}\quad If there is no $\thd{N-1}$ in $\calL$, then $\calL$ can be 
 trivially rewritten into the form $ \del_\tau(\ \ast\ ) + \tilde{Z}$
 and the lemma is proved. 
\parmedskipn
{\it Case 3:}\quad Now we come to the remaining case where 
there is one power of  $\thd{N-1}$ present in $\calL$. 
In this case, we have
\begin{eqnarray}
\calL &=& \thd{N-1}_\be \thd{N}_\ga Z_{\be;\ga} \comma 
\end{eqnarray}
and apriori there is no symmetry property required of $Z_{\be;\ga}$. 
Now look at the terms containing $\thd{2N-1}$ in $\calD_N[\calL,\theta_\al]$. 
Such terms can only be produced by $\del_\tau^N (\del \calL/\del
\thd{N}_\al)$ and $\del_\tau^{N-1} (\del \calL /\del \thd{N-1}_\al)$ and 
 never arise by acting on $Z_{\be;\ga}$. Explicitly, these two contributions
 add up to give 
\begin{eqnarray}
(-1)^{N-1}\thd{2N-1}_\be (Z_{\al;\be}+Z_{\be;\al})\period
\end{eqnarray}
For $N\ge 2$, obviously this cannot be proportional to $\thdot_\al$. 
It cannot contain $\Xi$ either, as was already shown in Lemma 1. 
Hence it must vanish and $Z_{\al;\be}$ has to be antisymmetric. 
In such a case, $\calL$ can be rewritten as 
\begin{eqnarray}
\calL &=& \del_\tau (\thd{N-1}_\be \thd{N-1}_\ga Z_{\be;\ga}) -\calL
 -\thd{N-1}_\be \thd{N-1}_\ga \del_\tau Z_{\be;\ga} \period
\end{eqnarray}
As in the previous two cases, we may solve for $\calL$ and  
the lemma follows. \qed
\subsection{Proof of the Proposition}
We are now ready to give a proof of the Proposition. 
\Proof Through the previous lemmas, the 
 problem of finding the solution to our basic equation has been 
 reduced to the case where $\calL$ contains $\theta$ and $\thdot$ only. 
The general form of $\calL$ is then 
\begin{eqnarray}
\calL &=& \sum  \theta_{\al_1}\theta_{\al_2}\cdots \theta_{\al_n}
\thdot_{\be_1}\thdot_{\be_2}\cdots \thdot_{\be_m} Z_{\al_1\al_2\ldots\al_n;
\be_1\be_2\ldots \be_m} \period
\end{eqnarray}
First consider the terms with more than two $\thdot$'s. Then from 
 $-\del_\tau  \del \calL /\del \thdot_\al$, we will get 
 a term  proportional to
\begin{eqnarray}
\theta_{\al_1}\theta_{\al_2}\cdots \theta_{\al_n} \thddot_{\be_2}
\thdot_{\be_3}\cdots \thdot_{\be_m}Z_{\al_1\al_2\ldots\al_n;
\al \be_2\ldots \be_m} \comma 
\end{eqnarray}
which cannot be produced in any other way. Such a term is allowed 
 if and only if it is proportional to $\thdot_\al$. For this to happen,
 $Z_{\al_1\al_2\ldots\al_n;
\al \be_2\ldots \be_m}$ must contain 
 $\delta_{\al\be_3}$ and similar Kronecker $\delta$'s. However, due to 
 the total antisymmetry of the index set $(\al, \be_2, \ldots, \be_m)$, 
 this is not possible. Hence such terms cannot be present in $\calL$. 

We are left with the possibility with zero or one $\thdot$ and 
 the most general form of $\calL$ is 
\begin{eqnarray}
\calL &=& \sum \left(\theta_{\al_1}\theta_{\al_2}\cdots \theta_{\al_n} 
Z_{\al_1\al_2\ldots
\al_n} +  \theta_{\al_1}\theta_{\al_2}\cdots \theta_{\al_n}
\thdot_{\be_1} Z_{\al_1\al_2\ldots
\al_n;\be_1}\right) \period
\end{eqnarray}
The terms with a $\thdot$ can be rewritten as
\begin{eqnarray}
\calL_1 &\equiv & \theta_{\al_1}\theta_{\al_2}\cdots \theta_{\al_n}
\thdot_{\be_1} Z_{\al_1\al_2\ldots
\al_n;\be_1} \nn\\
&=& \del_\tau (\theta_{\al_1}\theta_{\al_2}\cdots \theta_{\al_n}
\theta_{\be_1} Z_{\al_1\al_2\ldots
\al_n;\be_1}) \nn\\
&& -n \thdot_{\al_1} \theta_{\al_2}\cdots \theta_{\al_n} \theta_{\be_1}
Z_{\al_1\al_2\ldots
\al_n;\be_1} -\theta_{\al_1}\theta_{\al_2}\cdots \theta_{\al_n}
\theta_{\be_1} \dot{Z}_{\al_1\al_2\ldots
\al_n;\be_1}
\end{eqnarray}
Consider first the case where $n\ge 2$. Then if
 $Z_{\al_1\al_2\ldots
\al_n;\be_1} $ is not totally antisymmetric, the second and the 
 third term vanish and hence $\calL_1$ is a total derivative, which can be 
 dropped. On the other hand if $Z_{\al_1\al_2\ldots
\al_n;\be_1} $ is totally antisymmetric, then the second term 
 is equal to $-n \calL_1$. Thus we may solve for $\calL_1$ and get
\begin{eqnarray}
\calL_1 &=& {1\over n+1} \Biggl(\del_\tau (\theta_{\al_1}\theta_{\al_2}\cdots \theta_{\al_n}
\theta_{\be_1} Z_{\al_1\al_2\ldots
\al_n;\be_1}) \nn\\
&& -\theta_{\al_1}\theta_{\al_2}\cdots \theta_{\al_n}
\theta_{\be_1} \dot{Z}_{\al_1\al_2\ldots
\al_n;\be_1} \Biggr)
\end{eqnarray}
Thus, up to a total derivative, $\calL_1$ is reduced to an expression 
 without $\thdot$. 

The important exception occurs for  $n=1$, namely when
$\calL_1=\theta_\al \thdot_\be Z_{\al;\be}$. In this case there is no
 apriori symmetry properties required of $Z_{\al;\be}$. Rewriting this as 
 before, we get
\begin{eqnarray}
\theta_\al \thdot_\be Z_{\al;\be} &=& 
\del_\tau (\theta_\al \theta_\be Z_{\al;\be}) 
+\theta_\al \thdot_\be Z_{\be;\al} \period
\end{eqnarray}
If $Z_{\al;\be}$ is antisymmetric, then just as before $\theta_\al\thdot_\be
Z_{\al;\be}$ is a total derivative. On the other hand if $Z_{\al;\be}$ 
has a symmetric part, then such a term can survive without contradiction. 

At this point we have reduced the possible form of $\calL$ down to
\begin{eqnarray}
\calL &=& \sum \theta_{\al_1}\theta_{\al_2}\cdots \theta_{\al_n} 
Z_{\al_1\al_2\ldots
\al_n} + \theta Z_S \thdot \comma 
\end{eqnarray}
where $Z_S$ is a symmetric bispinor. Then, our basic equation becomes 
\begin{eqnarray}
 \sum n \theta_{\al_2}\cdots \theta_{\al_n} 
Z_{\al\al_2\ldots
\al_n} +2(Z_S\thdot)_\al  + (\dot{Z}_S \theta)_\al = \thdot_\al G \period
\end{eqnarray}
Look at the terms containing $\theta$ only. 
Since $Z_{\al\al_2\ldots\al_n}$ is totally antisymmetric whereas $\dot{Z}_S$ 
 is symmetric, these coefficient functions must vanish separately. 
In particular 
 it implies that $Z_S$ is a constant and, since the only 
 symmetric constant bispinor is $\delta_{\al\be}$, 
 it must be of the form  $(Z_S)_{\al\be} =c\delta_{\al\be}$, 
with $c$ a constant. Then the basic equation is satisfied with 
 $G=2c$. 

Thus, adding the $\theta$-independent contribution $\Ga[r]$ which has 
 hitherto been suppressed, we have found that 
the solution to the basic consistency equation 
 can only be of the form 
\begin{eqnarray}
\Ga &=& c\int d\tau (\theta\thdot)+\Ga[r] \period
\end{eqnarray}
Finally, requiring that the original Ward identity $\dzep \Ga=0$ be satisfied, 
$\Ga$ is fixed completely to be proportional to the tree-level 
 action. This completes the proof. \qed
\subsection{Proof of the Main Theorem}
Having established the Proposition, the proof of the Theorem
 follows immediately. 
\Proof Let $\delta$ be a quantum-corrected SUSY transformation operator 
 and let $\Ga$ be a solution to the SUSY Ward identity $\delta \Ga=0$. 
Now let us write the most general solution as 
$\Gatil=\Ga +\Delta \Ga$.
With $\Ga$ being a solution, $\Delta\Ga$ must by itself satisfy
 $\delta \Delta\Ga=0$. 
Since the tree-level solution is unique up to a constant, 
 we fix its normalization so that the tree-level 
 correction $\Delta\Ga^{(0)}$ vanishes. 
By assumption, we can expand $\delta$ and $\Delta\Ga$ in powers of
 the coupling constant $g$ as 
\begin{eqnarray}
\delta &=& \sum_{n\ge 0} g^{2n} \delta^{(n)}\comma \qquad 
 \Delta\Ga = \sum_{m\ge 1} g^{2m-2} \Delta\Ga^{(m)}\period
\end{eqnarray}
In our convention, $r_m, \theta, \tau$ and $g^2$ carries dimensions
 $1,{3\over 2}, -1$ and $3$ respectively and hence the dimension 
 of $\Delta\Ga^{(m)}$ is $3-3m$. 
The Ward identity for $\Delta\Ga$ then becomes, for each $k\ge 1$, 
\begin{eqnarray}
0 &=& \sum_{n=0}^{k-1} \delta^{(n)} \Delta \Ga^{k-n} \period
\end{eqnarray}
Explicitly, the first few equations are
\begin{eqnarray}
0 &=& \delta^{(0)}\Delta \Ga^{(1)} \comma \nn\\
0 &=& \delta^{(0)}\Delta \Ga^{(2)} + \delta^{(1)}\Delta \Ga^{(1)}\comma \\
0 &=& \delta^{(0)}\Delta \Ga^{(3)} + \delta^{(1)}\Delta \Ga^{(2)} 
+\delta^{(3)} \Delta\Ga^{(1)} \period \nn
\end{eqnarray}
Our Proposition dictates that the only possible solution to the first equation 
 is $\Delta\Ga^{(1)} = c\int d\tau (\dot{r}^2 +\theta\thdot)$, with 
 $c$ a constant independent of $g$. Such an expression does not have
 the right dimension and hence $\Delta\Ga^{(1)}$ must vanish. 
This in turn makes the second equation to be of the form 
 $\delta^{(0)}\Delta \Ga^{(2)}=0$ and for the same reason we get
 $\Delta \Ga^{(2)}=0$. Repetition of this reasoning clearly shows 
 that $\Delta \Ga=0$ altogether. This completes the proof. \qed
\section{Discussions}  
 \sectionnumbering
As the results of our present investigation have already been well-summarized,
 we shall not repeat them.  
 Below we shall discuss 
 two important future problems and indicate how they should be attacked.
\parsmallskip
One immediate question is whether similar results hold 
 for the multi-body case. Replacing $r_m$ and $\theta_\al$ by 
 $r_{m,i}$ and $\theta_{\al,i}$, where the extra index $i$ 
 refers to  different D-particles, it is easy to see
 that our consistency equation (\ref{basiceq}) is generalized 
 to 
\begin{eqnarray}
\sum_i (\thdot_{\ga,i} F_{\be,i} + \thdot_{\be,i} F_{\ga,i}) =0\period
\end{eqnarray}
Apart from the special case where $F_{\be,i}$ 
 contains a product of 16 $\thdot$'s for each
particle, this equation can be algebraically
 satisfied if and only if 
\begin{eqnarray}
F_{\be,i} &=& S_{ij} \thdot_{\be,j} \comma 
\end{eqnarray}
where $S_{ij}$ is a symmetric matrix.  There are many such matrices, 
examples of which are $r^m_i r^m_j$,  
$\theta_i \ga^{mn}\theta_j r_m \dot{r}_n$,  etc. Therefore, 
 the analysis of its solution becomes  more involved compared to
 the single D-particle case treated in this work. Nevertheless, 
 it would be quite interesting and important to see, first,
 if one can construct a non-trivial solution, 
thereby proving that SUSY alone cannot determine 
 the structure of the effective action uniquely for the 
 multi-body case.  If serious efforts in this 
direction should all fail, that would then indicate a possibility 
 to be able to construct a uniqueness proof.
\parsmallskip
The second and more ambitious enterprise is, as was  mentioned
 in the introduction,  to study if SUSY 
 alone truly determines the dynamics of the theory.  In other words, 
 one would like to know  whether 
 the requirement of SUSY determines {\it both} the effective 
transformation law {\it and} the effective action  at the same time, 
up to some  field redefinitions.  

To formulate this problem more precisely, 
 we must define what we mean by the effective SUSY transformation
 to begin with. In the absence of off-shell superfield formalism, 
 this is already non-trivial since the closure relations, only by which 
 one can guarantee that the transformation is SUSY, themselves 
must depend on the form of the full effective action. 
Consider, for example, the simplest case of the source-probe situation. 
Assuming that the non-canonical parts of the closure relations
are {\it linear} in the equations of motion\footnote{Although we 
 have no proof, judging from the results at tree and  1-loop 
 levels, non-linear dependence is highly unlikely.}, the set of equations 
characterizing SUSY would be of the form
\begin{eqnarray}
\delta_\ep \theta_\al &=& T_{\al\be}\ep_\be\comma  \\
\delta_\ep r_m &=& \Omega_{m\be}\ep_\be\comma  \\
\left[\delta_\ep, \delta_\lam\right] \theta_\al 
 &=& 2(\ep \lam) \thdot_\al + A_{\al\be\ga\delta}
{\delta \Ga \over \delta \theta_\delta}\ep_\be \lam_\ga 
+B_{\al\be\ga n} {\delta \Ga \over \delta r_n}\ep_\be \lam_\ga\comma  \\
\left[\delta_\ep, \delta_\lam\right] r_m
 &=& 2(\ep \lam) \dot{r}_m + C_{m\be\ga\delta}
{\delta \Ga \over \delta \theta_\delta}\ep_\be \lam_\ga 
+D_{m\be\ga n} {\delta \Ga \over \delta r_n}\ep_\be \lam_\ga \comma \\
\delta_\ep \Ga &=& 0 \comma 
\end{eqnarray}
where $\ep_\be$ and $\lam_\ga$ are the global SUSY parameters and $A\sim D$ 
 are some, in general field dependent, coefficients. 
To show that SUSY determines the dynamics completely, one must solve 
 this set of non-linear functional differential equations, with 
 the tree-level information as the only input,  and demonstrate 
 that all the quantities are fixed
uniquely,  up to field redefinitions. A preliminary investigation
 indicates that at least at low orders this might be feasible. 

We hope to report on these and related matters in future communications. 
\par\bigskip\noindent
{\large\bf Acknowledgment}\par\smallskip\noindent
Y.K. is grateful to T. Yoneya for a useful discussion. 
The research of Y.K. is supported in part by 
Grant-in-Aid for Scientific Research
on Priority Area \#707 \lq\lq Supersymmetry and Unified Theory of 
 Elementary Particles" No.~10209204 and 
 Grant-in-Aid for Scientific Research (B) 
No.~12440060, while that of T.M. is supported in part by 
the Japan Society for Promotion of Science under the Predoctoral
Research Program No.~12-9617, 
all from the Japan Ministry of Education, Culture, Sports, Science
 and Technology. 


\begin{thebibliography}
\bibitem{bfss} T. Banks, W. Fischler, S. H. Shenker and L. Susskind,
Phys. Rev. {\bf D55} (1997) 5112, hep-th/9610043.
\bibitem{susskind} L. Susskind, hep-th/9704080.
\bibitem{Pabanetal1} S. Paban, S. Sethi and M. Stern,
Nucl. Phys. {\bf B534} (1998) 137, hep-th/9805018.
\bibitem{Pabanetal2} S. Paban, S. Sethi and M. Stern,
J. High Energy Phys. {\bf 06} (1998) 012, hep-th/9806028.
\bibitem{lowe} D. A. Lowe, J. High Energy Phys. {\bf 11} (1998) 009, 
hep-th/9810075.
\bibitem{ss} S. Sethi and M. Stern,
J. High Energy Phys. {\bf 06} (1999) 004, hep-th/9903049.
\bibitem{Hyunetal} S. Hyun, Y. Kiem and H. Shin, Nucl. Phys. 
{\bf B558} (1999) 349, hep-th/9903022.
\bibitem{np} H. Nicolai and J. Plefka, 
Phys.\ Lett.\ B {\bf 477}, 309 (2000), hep-th/0001106.
\bibitem{Kaz-Mura1} Y. Kazama and T. Muramatsu, Nucl. Phys. {\bf 584}
 (2000) 171, hep-th/0003161.
\bibitem{Kaz-Mura3} Y. Kazama and T. Muramatsu, hep-th/0106218.
\bibitem{BBPT}
K.~Becker, M.~Becker, J.~Polchinski and A.~Tseytlin,
Phys.\ Rev.\ D {\bf 56}, 3174 (1997), hep-th/9706072.
\bibitem{Okawa-Yoneya} 
Y. Okawa and T. Yoneya, Nucl. Phys. {\bf B538} (1999) 67,
 hep-th/9806108.
\bibitem{Okawa-Yoneya2} Y. Okawa and T. Yoneya, Nucl. Phys. {\bf B541} 
(1999) 163, hep-th/9808188.
\bibitem{TR} W. Taylor and M. V. Raamsdonk, J. High Energy
	Phys. {\bf 04} (1999) 013, hep-th/9812239.  
\bibitem{Kaz-Mura2} Y. Kazama and T. Muramatsu, Class.\ Quant.\ Grav.\
	{\bf 18}, 2277 (2001), hep-th/0103116.
\end{thebibliography}
\end{document}